# Channel Performance Estimations with Extended Channel Probing

Kaida Kaeval
*Advanced Technology*
*ADVA Optical Networking SE*
Martinsried, Germany
kkaeval@adva.com

Helmut Grieβer
*Advanced Technology*
*ADVA Optical Networking SE*
Martinsried, Germany
hgriesser@adva.com

Klaus Grobe
*Global Sustainability*
*ADVA Optical Networking SE*
Martinsried, Germany
kgrobe@adva.com

Jörg-Peter Elbers
*Advanced Technology*
*ADVA Optical Networking SE*
Martinsried, Germany
jpelbers@adva.com

Marko Tikas
*Transmission Networks*
*Tele2 Estonia AS*
Tallinn, Estonia
marko.tikas@tele2.com

Gert Jervan
*Department of Computer Systems*
*Tallinn University of Technology*
Tallinn, Estonia
gert.jervan@taltech.ee

***Abstract*— We test the concept of extended channel probing in a Optical Spectrum as a Service scenario in coherent optimized flex-grid long-haul and 10Gbit/s OOK optimized 100-GHz fixed-grid dispersion managed legacy DWDM networks. An estimation accuracy better than ±0.1dB in GSNR implementation margin is obtained for both networks by using flexible coherent transceivers on lightpaths up to 822km on regional haul legacy network and up to 5738km on a long-haul network. We also explain, how to detect the operation regime of the channel by analyzing the received GSNR results from constant power spectral density (PSD) probing mode and constant signal power probing mode.**

## I. INTRODUCTION

While coherent optimized DWDM networks are becoming the standard for long haul communication networks, many access and regional haul data transmission networks still rely on dispersion managed 10-Gbit/s OOK (on-off keying) optimized networks. While a lot of effort is put into upgrading these networks, full implementation of the decisions may take years and intermediate solutions are required by the operators to satisfy the growing need for capacity. One such solution is to enable coherent transmission over the legacy DWDM networks.

Considering the operation with design based signal power levels and sufficient guard bands to avoid interference with the existing 10Gbit/s OOK signals, possible filtering penalties from various multiplexers, Reconfigurable Optical Add Drop Multiplexers (ROADMs) or grating based dispersion compensation (DCG) modules may still complicate the selection of the best possible transponder configuration from thousands of available configurations provided by the modern flexible symbol rate and modulation format transponders to achieve the highest capacity per channel.

Although the Gaussian Noise model based channel probing method [1],[2] to estimate the channel performance is primarily performed over uncompensated coherent optimized networks, where several studies have shown successful results[2],[3],[4],[5], the concept is useful also in estimating the performance in dispersion compensated networks [7]. In this paper, we show how extended channel probing allows to improve the prediction accuracy from previously reported 0.7dB [5] to less than 0.1dB accuracy for both flex-grid uncompensated long-haul and fixed-grid dispersion compensated regional-haul networks in order to fully utilize the flexibility of the coherent transponders. We discuss the PSD based probing mode and propose an operation regime analysis to detect operation in the nonlinear regime.

This work has received funding from the German ministry of education and research (BMBF) under grant No. 16KIS0989K (OptiCON project).

## II. EXTENDED CHANNEL PROBING

In order to evaluate the actual channel performance of a lightpath at hand, characterized probing-light is inserted into the network in the corresponding channel location and the bit error ratio (BER) estimation of the receiver, converted into a Q-value is used to estimate the respective effective Generalized OSNR (GOSNR). This value is then normalized to the symbol rate of the Probing Light Transceiver (PLT) signal to obtain the GSNR. However, the received GSNR value here would only be applicable to describe the spectrum performance for the exact covered bandwidth by the PLT configuration at the exact probed power regime.

To overcome this and get more insights to the performance of the network and spectral slot at hand, a set of PLT configurations with different modulation formats and symbol rates can be used. For the current work, eleven PLT configurations as listed in Tab. 1 were characterized by a back-to-back measurement of the Q-factor versus the OSNR in the lab. The logarithmic Q-over-OSNR characteristic was then fitted by a 2nd order polynomial to estimate the Generalized OSNR (GOSNR) from the Q-factor of each PLT configuration. This value was then normalized to the symbol rate of the PLT signal to obtain the GSNR.

Based on the estimated GSNR values from eleven PLT configurations, we calculate the average estimated GSNR based on the PLT configuration estimations, that do not experience severe penalties from filtering, making this method usable also in systems using fixed filter access or other filtering elements.

Finally, the GSNR implementation margin is estimated for a pool of usable transponder configurations, leaving out all configurations with higher symbol rate than the PLT with highest symbol rate configuration still working over the link. To calculate the GSNR implementation margin, we subtract the typically required GSNR margin according to the system specification from the estimated GSNR [5]. All calculations resulting in positive GSNR implementation margins are expected to work over the probed link.

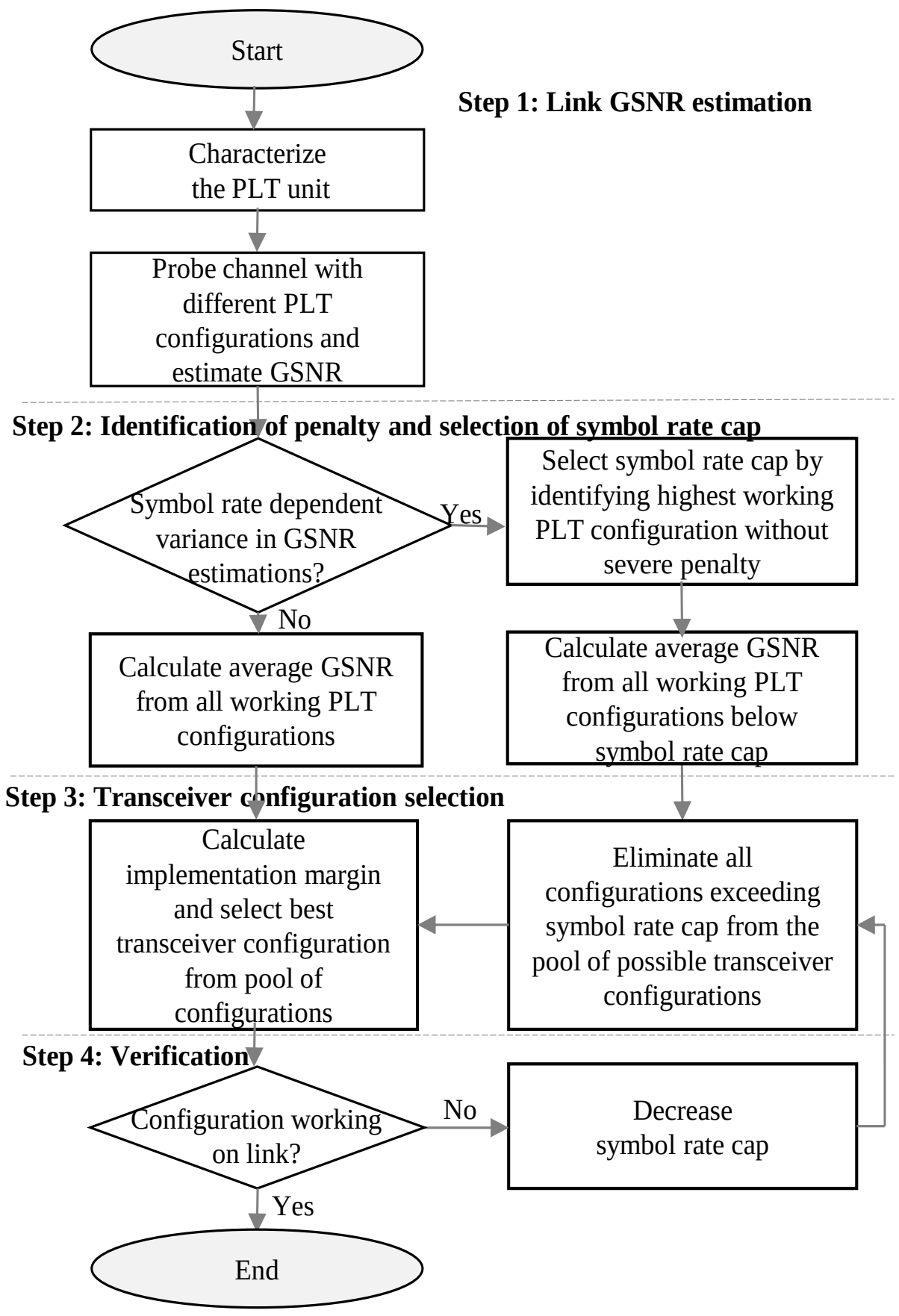

**Fig. 1:** Extended channel probing flowchart

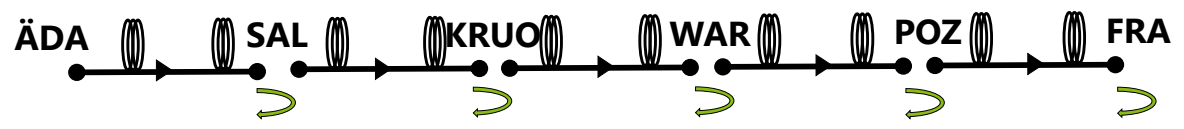

| Loopback location | DCM data | Looped link length(km) | Spans (back and forth) |
|---|---|---|---|
| LH SAL | DCM free | 1016 | 14 |
| LH KRUO | DCM free | 1792 | 24 |
| LH WAR | DCM free | 2943 | 36 |
| LH POZ | DCM free | 3751 | 48 |
| LH FRA | DCM free | 5738 | 74 |

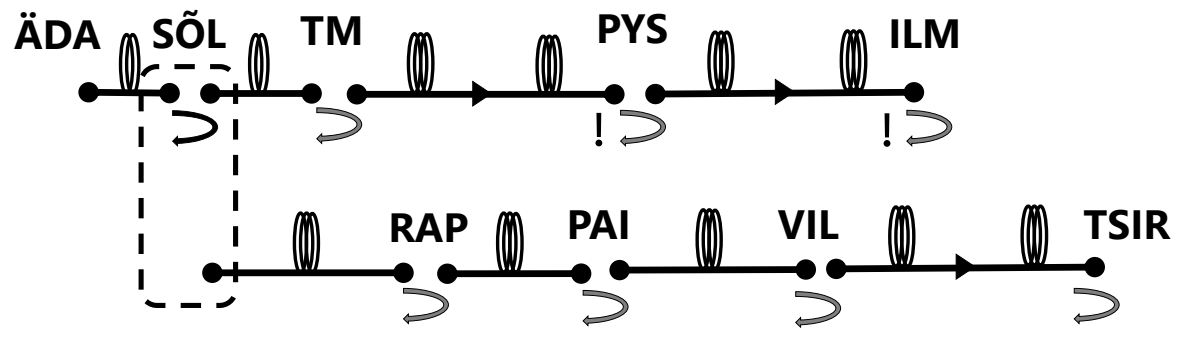

| Loopback location | DCM data | Looped link length(km) | Spans (back and forth) |
|---|---|---|---|
| R SOL | DCM free | 3 | 2 |
| R TM | DCF only | 70 | 4 |
| R RAP | DCF only | 144 | 4 |
| R PAI | DCF only | 241 | 6 |
| R VIL | DCF only | 382 | 8 |
| R TSIR | DCF only | 675 | 12 |
| R PYS | DCF+DCG Mix | 485 | 8 |
| R ILM | DCF+DCG Mix | 822 | 12 |

**Fig. 2:** Network links tested

By using the same probe configurations as verification signals, comparison of the estimated GSNR implementation margin to the actual performance of the verification signals was performed.

## III. Probing Mode

To start with extended channel probing, a switch from constant signal power setting to constant Power Spectral Density (PSD) power setting has to be performed. Although not as intuitive to work with, it has multiple benefits. Most important are the good alignment with power limited networking caused by the amplifier total output power and better comparability of the received GSNR results.

In case of a constant power distribution over the full C-band, the maximum usable PSD is defined by the maximum total power to the line from the amplifier output and total C-band bandwidth. As these values might have been optimized per link or hop, calculations for the suitable PSD in the production networks can be performed according to the ROADM output power-levels. When calculating the PSD for fixed-grid 10Gbit/s optimized legacy networks, the original channel power to the line can be divided with the 10Gbit/s bandwidth. That would also ensure the optimum performance with dispersion compensation fibres (DCF).

Network equalization should be performed with signal configuration with the highest symbol rate, regardless of possible penalties this signal format may experience due to filtering. The motivation here is to set the internal variable optical attenuators in the ROADM modules to a level that ensures that symbol rate changes during the probing activity would not increase the total power levels per open optical cross-connect in the network. In our tests, we used 69GBd signals providing the highest power per channel.

As a consequence, the PSD based probing mode allows better comparability of the estimated GSNR values of each PLT configuration, as equal GSNR is expected at the receiver for the signal configurations with equal PSD in near-linear operation regime, where differences in results are mainly caused by transceiver implementation penalties, measurement errors or network fluctuations.

## IV. Test Set-up and Infrastructure

The concept of extended channel probing shown in Fig. 1 was tested over two different production networks. A coherent optimized flex-grid long-haul network to verify the method and selected PSD based probing mode and a 10-Gbit/s optimized 100GHz-grid legacy regional-haul network with dispersion compensation modules utilizing both, DCF and DCG modules. Network diagrams and link data is presented in Fig. 2, where links with exclamation mark loopbacks also included DCG modules.

**TABLE. I:** PLT configurations

| Line rate | Modulation | Symbol rate |
|---|---|---|
| 100Gbit/s | DP-QPSK | 31.5GBd |
| 200Gbit/s | DP-16QAM | 34.7GBd |
| 300Gbit/s | DP-64QAM | 34.7GBd |
| 300Gbit/s | DP-32QAM | 41.7GBd |
| 200Gbit/s | DP-8QAM | 46.3GBd |
| 400Gbit/s | DP-64QAM | 46.3GBd |
| 300Gbit/s | DP-16QAM | 52.1GBd |
| 400Gbit/s | DP-32QAM | 55.6GBd |
| 200Gbit/s | DP-QPSK | 69.4GBd |
| 300Gbit/s | DP-8QAM | 69.4GBd |
| 400Gbit/s | DP-16QAM | 69.4GBd |

First, a total of five links with link lengths from 1016km to 5738km in a long-haul pan-European coherent optimized flex-grid network were tested to verify the methods capability to return expected results in a network well suitable for the channel probing. Then, eight links with link lengths from 3km to 822km were tested in a regional haul DCM-based 10Gbit/s OOK optimized network.

Both networks utilize a different set of infrastructure components, like multiplexers, amplifiers, boosters and ROADMs, but are built on standard single mode fibres (SSMF ITU-T 652.D). This sets a good comparison for demonstrating the capabilities of extended channel probing for estimating additional aspects of the link performance at an Optical Spectrum as a Service scenario independent of the network infrastructure.

## V. Overall Result Accuracy

To better visualize the differences in probing in legacy networks and coherent optimized flexgrid networks, in Fig. 3, raw estimated GSNR data from the extended channel probing in both networks is shown. Markers with different styles refer to configurations with different symbol rates.

As visible in the inset of Fig. 3, the results are well aligned for the long haul network. Each PLT is contributing to the overall accuracy of the estimation and the average GSNR estimation over all PLT configurations allows us to reduce the error in GSNR implementation margin from previously reported 0.7dB [5] to 0.05dB. The reason is that, while small fluctuations in the network cause different BER readings, and thus the Q-value per each measurement also varies, the extended approach allows us to average out the variance in the results and get a better estimate for the link GSNR. GSNR implementation margins calculated from averaging GSNR values, together with the actual performance indication for the verification signals, are presented in Fig. 4. Different marker styles refer to different symbol rate regimes and line styles distinguish different modulation formats. The chart is zoomed in for the area near zero implementation margin and check or fail marks indicate, if the configuration worked in reality or not.

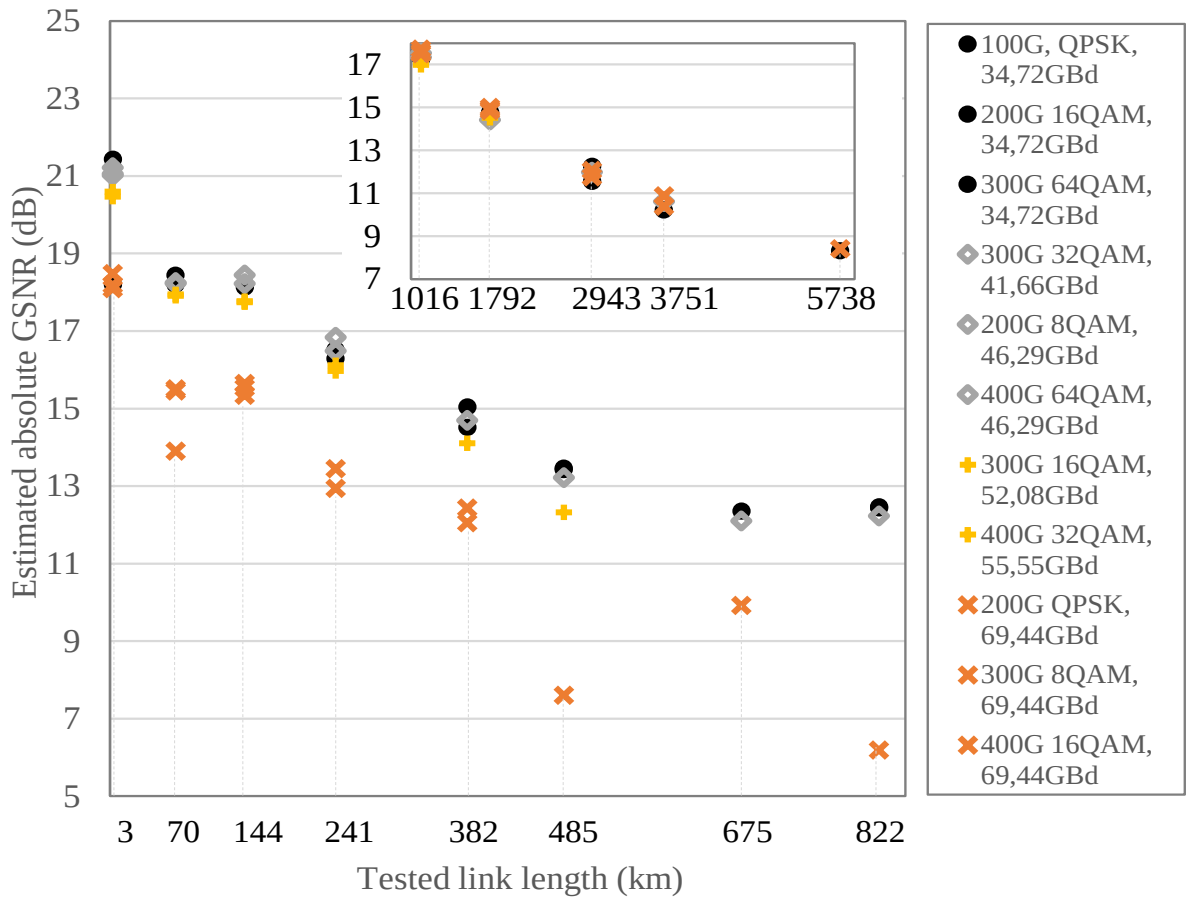

**Fig. 3:** Raw data from extended channel probing in regional haul network, estimated by different probe configurations. Inset presents the extended probing results for long-haul network.

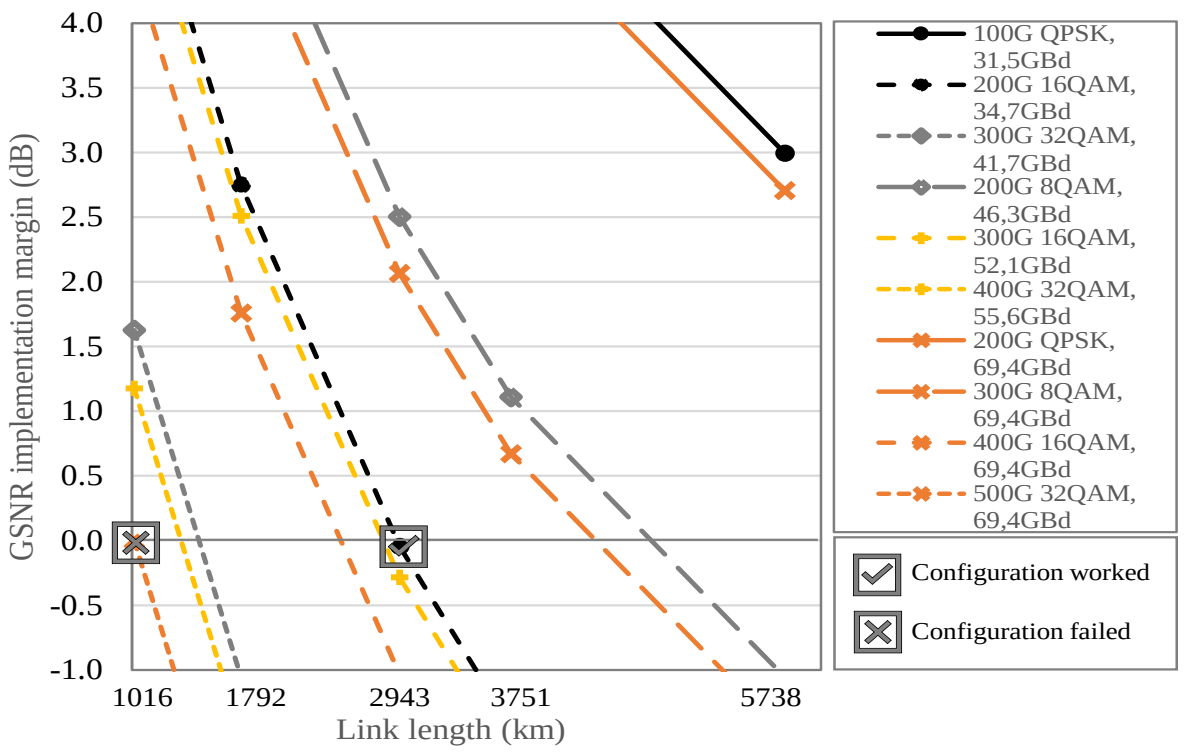

**Fig. 4:** GSNR implementation margin compared to channel actual performance in a long-haul network.

In contrast, for the legacy regional haul network it is relatively difficult to make any assumptions on the actual performance capabilities due to wide spread of the estimated GSNR values, reducing the average estimated GSNR value just as a statistical number. In this case, further analyses on the raw data are required to be able to decide the best performing modulation and symbol rate combination for the current spectral slot at hand.

## VI. Determining Highest Usable Symbol Rate

As high symbol rate signals are subject to stronger narrow band filtering, using a wideband probing signal results in underestimating the link GSNR that would be achievable for lower symbol rate signals. Using narrow band probing configurations would create the illusion of a link with a high GSNR not achievable for higher symbol rate signals.

In order to estimate the best possible modulation format and symbol rate combination in 10-Gbit/s optimized legacy networks, a bandwidth limitation caused by filtering penalty has to be identified.

According to our test results, when operated with system design based PSD powers, PLT configurations with equal symbol rate experienced similar penalties from the system and estimated the link GSNR with +/-0.35dB accuracy, regardless of the modulation type for up to 60GBd signals, whereas a change in the PLT symbol rate resulted in a great variance in the estimated GSNR.

From Fig. 3, we can see, that while the maximum difference between the estimated GSNR is below 0.7dB for the probe settings between 31.5 and 46.3GBd and below 1.1dB for the probes up to 55.6GBd, it quickly grows up to 7.1dB for the probes including 69.4GBd. This behaviour can be explained with a filtering penalty from the channel filter modules (3dB bandwidth of 80 GHz) and ROADMs for the DCF based links and additional penalty from DCG modules (3dB bandwidth of 60 GHz) for the 485-km and 822-km links.

To identify the bandwidth limitation and select the highest usable symbol rate on the link, we compare the estimated absolute GSNR values from the different probes and select the working probe with the highest symbol rate configuration that does not experience severe penalties from filtering. This symbol rate value (symbol rate cap) is essential in determining the best working modulation format and symbol rate configuration for the current link. Looking at Fig. 3, this means leaving out all high symbol rate configurations marked with "x" for the links with channel filter modules or with "+" for even smaller bandwidth due to DCG modules. Then, the average estimated GSNR can be calculated for the link as described for the long-haul network scenario and the implementation margin is calculated for all configurations with smaller and equal symbol rate than the highest working PLT configuration. Finally, the best transceiver configuration can be selected as per highest line rate from all possible configurations returning positive implementation margin[5]. Additional margins can be added to the GSNR estimations to increase the robustness of the prediction to any fluctuations and ageing effects in the network[6].

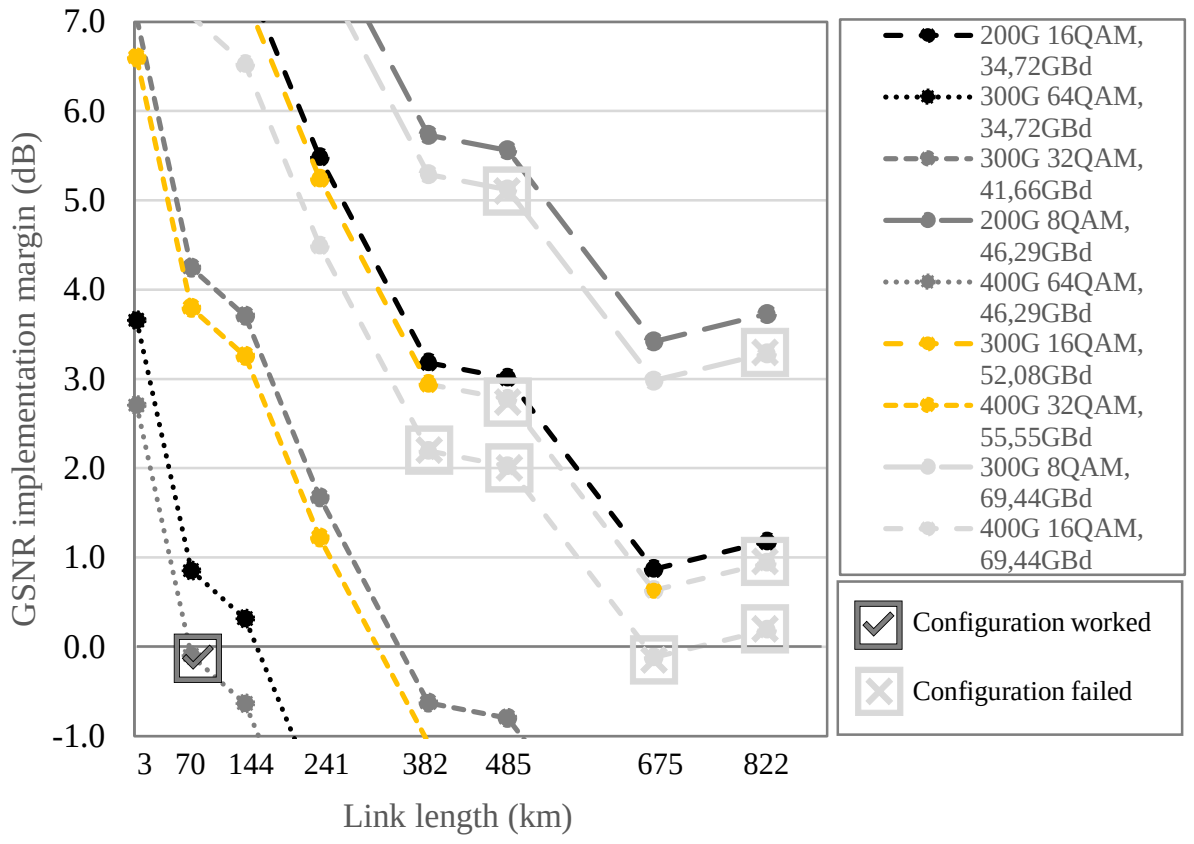

**Fig. 5:** GSNR implementation margin compared to channel actual performance in a regional-haul network.

The results for the extended channel probing method with additional step considering the bandwidth limitation and symbol rate cap for the pool of possible transponder configurations are presented in Fig. 5. Similarly to Fig. 4, different marker styles refer to different symbol rates and line styles distinguish modulation formats. High symbol rate configurations not usable on the link and therefore left out from the GSNR implementation margin calculations are presented as light grey lines. The chart is zoomed in between -1 and +7dB GSNR implementation margin, to better illustrate the importance of the symbol rate limitation on the selection of the best working transceiver configuration. Similarly to Fig. 4, check or fail marks indicate if the configuration in reality worked or not. Light grey fail marks indicate the failing configurations exceeding the bandwidth limitations/symbol rate cap that were expected not to work.

Following the extended channel probing method, we were able to achieve the selection of the best possible modulation format and symbol rate combination on any link with an accuracy better than 0.1dB of GSNR implementation margin also in the dispersion managed legacy network. Considering the margin values usually implemented for network robustness against aging, slow changes in the network and infrastructure, the achieved result is an accurate measure of the network performance at a current moment.

## VII. Detecting Operation Regime

In addition to the accurate GSNR estimations, extended channel probing allows to evaluate the operation regime of the channel. To this end, results received with constant PSD over different symbol rates are compared to the constant signal power measurements.

For the network equalized with the highest symbol rate signal, providing the highest to the line power as per network design, this initial power can be taken as a reference point (reference point 1 in Fig. 6) for the operation regime test, regardless of the filtering penalty it experienced.

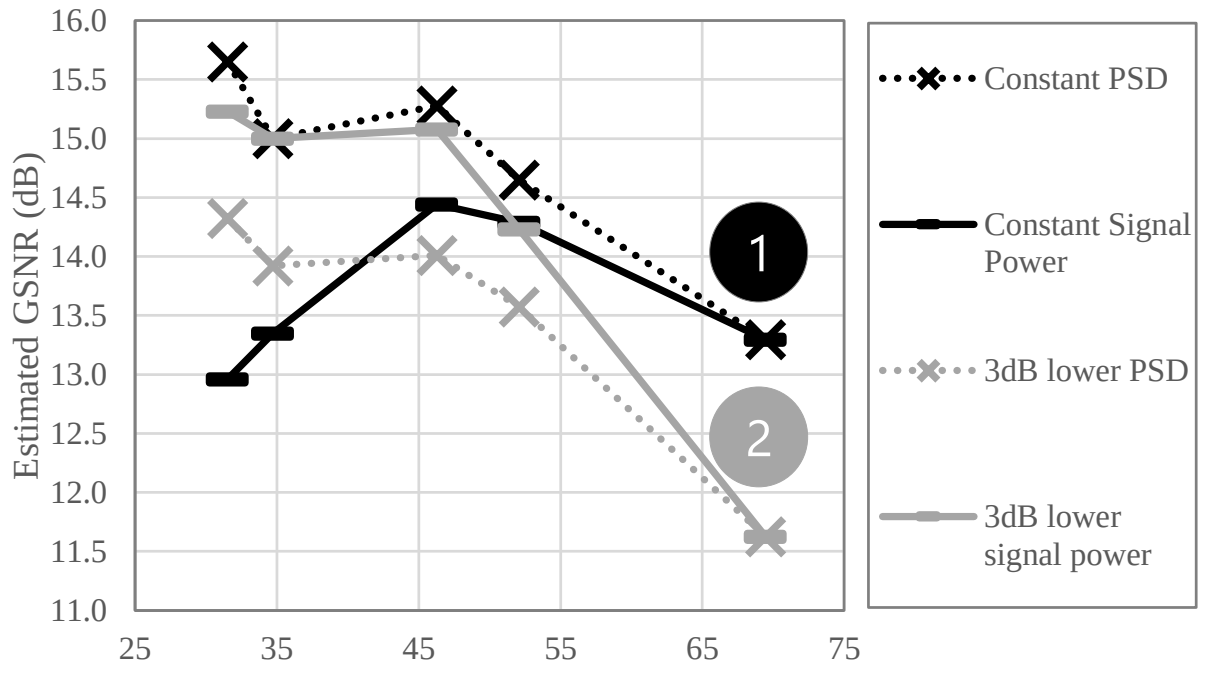

**Fig. 6:** Operation regime estimation

The 382-km link from our tests was analysed by comparing design based PSD mode GSNR estimations to design based signal power mode GSNR estimations. In both cases, the total power per probing activity stays equal or below the reference power used for the network equalisation. By comparing the estimated GSNR values for a black dotted, constant PSD line and a black solid, constant signal power line, we can see that signals perform better at constant PSD. This indicates that, with constant signal power, the increased PSD of individual narrowband signals is too high and the signal power is above the optimum balance between ASE noise and nonlinear distortions. For example, if we are operating in a linear regime, corresponding to a 3dB lower reference point signal power (reference point 2) the lines swap positions for narrower channels, and equal PSD delivers lower GSNR estimations than constant signal power mode. This means operation in the linear regime, as signals still do benefit from increased total signal powers.

This kind of analysis can be useful especially for longer links to adjust signal powers in the right direction to improve the performance for signal configurations operating at the BER threshold.

## VIII. Conclusions

We use extended channel probing to improve the accuracy of the link GSNR estimations and to select the best possible transceiver configuration for coherent transponders operated on both, flex-grid coherent optimized and 10-Gbit/s optimized fixed-grid dispersion managed legacy DWDM systems. We show that eleven characterized probe configurations are sufficient to reliably select the best modulation format and symbol rate configuration for the transceivers and also derive useful information for network development and operation decisions.

In future studies, we want to investigate the minimum set of PLT configurations that can lead to a similar estimation accuracy.

## Acknowledgment

We would like to thank Tele2 Estonia for their continuous co-operation and help regarding the research on Optical Spectrum as a Service in Disaggregated Networks.